**Fabrication and Characterization of Boron Carbide / n-Silicon Carbide Heterojunction Diodes**

S. Adenwalla(1) , P. Welsch, A. Harken, J. I. Brand(2), A. Sezer(2), and B. W. Robertson

Departments of Mechanical Engineering, (1)Physics and Astronomy, and (2)Chemical Engineering, and the Center for Materials Research and Analysis, University of Nebraska-Lincoln, Lincoln, NE 68588, U.S.A.

The fabrication, initial structural characterization and diode measurements are reported for the first boron carbide / silicon carbide heterojunction diode. Current-voltage curves obtained for operation at temperatures from 24 °C to 388 °C. PECVD-deposited undoped boron carbide material is highly crystalline and consists of a variety of polytypes of B-C with crystal sizes as large as 110 nm. Crystal phases are similar to those for PECVD B-C on Si but only partially match known boron and boron-rich B-C phases.

Boron carbide has long been valued as a hard, refractory, corrosion-resistant ceramic and investigated as a potential electronic material [1]. If combined with useful semiconducting properties, it mechanical properties male boron carbide attractive for devices that may function reliably at high temperatures in harsh chemical environments. However, only in the past ten years has a semiconducting form of boron-carbide been successfully grown, by plasma enhanced chemical vapor deposition (PECVD), and used in a variety of initial diode and transistor devices. [2-4] The boron-rich boron-carbon compounds share basic structural similarities. The building block of these structures is the $B_{12}$ icosahedron of rhombohedral boron. At the carbon rich end of the wide composition range for the equilibrium phase designated as $B_4C$, there may be intericosahedral chains of boron and carbon in the crystal unit cell, as well as replacement of some icosahedral boron atoms by carbon atoms. The structural similarities of these compounds imply that the energy barriers between these different structures may be small and may lead, under non-equilibrium conditions, to simultaneous formation of a number of different polytypes, as occurs in silicon carbide, for example. Not all boron-rich B-C crystals are semiconducting. The crystallinity of semiconducting B-C films grown by PECVD has been questioned despite the Bragg diffraction peaks in published x-ray diffraction patterns.

Here we report on the first fabrication of boron carbide / silicon carbide heterojunction diodes and on their initial electrical and crystalline structure characterization.

Boron carbide was grown on SiC substrates by plasma enhanced chemical vapor deposition (PECVD) using a single precursor, orthocarborane – *closo*-1,2-dicarbadodecaborane ($C_2B_{10}H_{12}$) – that has a boron to carbon ratio that lies in the solid solution composition range of $B_4C$. Substrates were cleaved to suitable sizes from as-received n-type (111) 6H-SiC wafers (SiCrystal AG, Germany) and cleaned immediately before insertion into a custom-designed parallel-plate radio frequency plasma reactor. A substrate heater temperature of 350 °C was maintained throughout initial plasma etching (at 200 mTorr and 10 sccm Ar for 30 minutes) and subsequent deposition. Orthocarborane was sublimed at ~ 50 to 70 °C and introduced through a heated manifold by a flow of argon carrier gas. Depositions, using ~25 W of 13.56 MHz RF power for 15 minutes at ~ 200 mTorr reactor chamber pressure, were followed by cooling over one hour to 250 °C. Coated substrates were then cooled to room temperature over about one additional hour and were stored in vacuum until used.

Gold contacts, about 1.5mm in diameter and at least 100nm thick, were sputtered on the B-C surface and on the adjacent SiC surface, on the same wafer face as the B-C deposits in order to minimize the influence of SiC micropipes on the electrical performance. Robust external connections were provided through electrically conductive epoxy pads on top of the gold contacts. Devices were heated on a hot plate while current and voltage measurements were obtained using a Keithley 2400-C SourceMeter and TestPoint software.

The diode current-voltage (I-V) curves in Figure 1, obtained from the Au/B-C/n-SiC/Au device for a range of temperatures from 24°C to 388 °, are the first reported data for such a device. The SiC device displays much better diode characteristics than those grown on Si, although the present data were obtained from an unencapsulated device in laboratory air. The reverse saturation currents are lower and the breakdown voltages are higher than for their Si counterparts. Although the B-C/n-SiC device clearly retains diode characteristics to 388 °C, there is a deviation in behavior with increasing temperature, particularly of the forward current dependence on applied bias at 388°C. (This deviation may relate to the lack of encapsulation, rather than to any other source.)

From fitting of the linear portions of logarithmic plots of current as a function of applied bias, ideality factors and series resistances were determined. The ideality factors, decreasing from

~20 to under 2 with increasing temperature, are comparable with those for diodes based on microcrystalline Si. The dependence of the derived series resistances $R_s$ on the inverse absolute temperature (inset in Figure 1) is quite linear. The series resistance consists of the contact resistance and the resistance of the B-C layer, as well as the effects of carrier concentration. Typically, the concentration of holes (the majority charge carriers in undoped boron carbide) has an Arrhenius-like temperature behavior and the associated energy barrier is the difference between the pseudo-Fermi level and the energy band for the holes. Assuming that temperature mainly affects the concentration of carriers, but not their mobilities, and that only majority charge carriers need to be considered, the slope of a best-fit straight line to the series resistance data gives a characteristic energy barrier for the concentration of holes. For the sample grown on n-SiC, this characteristic energy barrier is $E = 0.11$ eV, whereas for the sample grown on n-Si it is much larger, at $E = 0.35$ eV. These energies are comparable with the activation energies previously found for states pinned near the conduction band edge in PECVD B-C. [5, 6]

Understanding and applying the electrical behavior of semiconducting B-C in electronic devices demands knowledge of the B-C phases that are present and of their properties. We have therefore conducted initial structural characterization, as follows.

X-ray diffraction angle measurements were acquired using a Rigaku theta – two theta x-ray diffractometer with Cu-Kα radiation. PECVD B-C deposited on SiC shows sharp, well-defined peaks (Figure 2a). The peak widths indicate that the B-C material has relatively large nanocrystallite dimensions (~ 30 to 110 nm). The peak positions, however, do not correspond to only one known boron carbide crystal type.

Figures 2b and 2c contain the data for samples grown on Si under similar conditions at different times. The peak positions and widths are identical within measurement error and only the relative peak heights are different. This confirms that the growth process is highly reproducible and that significant differences occur only in the orientations of crystals or in the proportions of polytypes or in both orientations and proportions. The peak widths correspond to crystals that are somewhat smaller (~ 10 to 100 nm) than those grown on SiC, indicating that crystalline growth is somewhat better on SiC than on Si.

By comparison of the X-ray diffraction pattern for growth on SiC (Figure 2a) with those for growth on Si (Figures 2b and 2c), ignoring the many substrate peaks, it is clear that the choice of substrate influences the growth, although not substantially under the present conditions. There are a number of peaks marked ■ in the figure that originate uniquely from the B-C layer and that match in position to within ~ 0.2° of each other for both Si and SiC substrates. (This corresponds to a match within 1.4% to 0.3% in interplanar spacings from peaks at 2-θ Bragg diffraction angles 14.4° to 78°, respectively.)

The structures of known boron carbide compounds closely resemble each other and elemental boron. Elemental rhombohedral boron consists of icosahedra of boron atoms placed at the corners of a rhombohedral unit cell. The addition of carbon leads to the formation of a chain of B and C atoms on the body diagonal of the unit cell and / or substitution of some B in the icosahedra by carbon. There may be small changes in the dimensions and structure of the unit cell as carbon is added to the compound, but the resulting effects on x-ray diffraction are subtle – often altering the main diffracting angles by small fractions of a degree. Distinguishing between the various boron carbide compounds in a micro- or nanocrystalline material using X-ray diffraction is therefore difficult, particularly since boron and carbon differ by one in atomic number and so their X-ray scattering lengths are also quite comparable.

Each molecule of orthocarborane consists of a $B_{10}C_2$ icosahedral core surrounded by 12 hydrogen atoms that are readily removed in the PECVD process. FTIR measurements on previously deposited thin films show very little evidence of C-H or B-H bonding and atomic absorption tests of previous samples show less then 5% hydrogen. Hence, we restricted our x-ray diffraction peak search to compounds containing only boron and/or carbon, but not hydrogen.

The diffraction peaks originating solely from the B-C layer and not, even partially, from the substrates are given in Table I. The peak at 14.24° is far below the first peaks for rhombohedral boron and its similarly structured B-C counterparts, which occur close to 19°. The crystal unit cell that yields the 14.24° peak is therefore much larger than the rhombohedral cell of these structures and corresponds closely with tetragonal boron or with the I-tetragonal boride $B_{48}(B_2C_2)$ structure first measured by Will and Ploog [7] by electron diffraction. The unit cell in $B_{48}(B_2C_2)$ consists of four boron icosahedra in a tetragonal arrangement, like that in tetragonal boron, but with the two carbon atoms and the two remaining boron atoms occupying intericosahedral sites. The other peaks are associated with interplanar spacings that correspond closely with some of those found in $B_{48}(B_2C_2)$, tetragonal B, rhombohedral B, $B_{12}(BC_2)$, or $B_{41.1}C_{4.45}$ but are not consistent with only one compound. For each of these possible phases, the intensities within each group of peaks do not scale according to the intensities for randomly oriented powders given in the International Crystal Diffraction Database Powder Diffraction Files – thus indicating at least some degree of preferred crystalline orientation. In-plane x-ray diffraction and cross-sectional TEM measurements will elucidate the degree of the texture. Initial high-resolution electron microscopy of as-deposited B-C previously grown on Si under slightly different conditions has revealed 3 nm-sized crystals. Lattice fringes in these crystals correspond to interplanar spacings $0.21 \pm 0.01$ nm, $0.24 \pm 0.01$ nm and $0.32 \pm 0.01$ nm [8] that are close to the present x-ray diffraction results. TEM is being used to determine the specific phases and their distributions more exactly.

In summary, undoped-B-C / n-SiC devices show good semiconductor diode behavior, at least from room temperature to 388 °C, that is superior to the B-C / n-Si behavior previously reported. The superior diode performance of the device on SiC may relate to its improved crystallinity. The structure of PECVD B-C depends on the deposition conditions and, somewhat, on the substrate. Under present PECVD reactor conditions, nanocrystalline material is deposited that corresponds closely to several boron and boron-carbon phases that we have identified but not fully to any one of these phases.

It is clear that the preliminary electrical assessment is only an indicator that high temperature operation may be quite successful. It is important now to identify the phase that contributes the most desirable semiconducting behavior in order that this promising semiconducting system can be better understood and applied in high temperature and other areas.

This work was supported by the Air Force Office of Scientific Research, the National Science Foundation, and the Nebraska Research Initiative. X-ray diffraction and electron microscopy used the central facilities of the Center for Materials Research and Analysis at the University of Nebraska-Lincoln. BR is grateful to Professor John Titchmarsh and the Department of Materials at the University of Oxford, England, for the invitation to conduct research on sabbatical in Oxford and for the provision of access to electron microscopy facilities there.

**Figure Captions**

Fig. 1: Current – applied bias (I-V) curves of an Au / B-C / n-SiC / Au device as a function of temperature. Curve labels are temperatures in °C. Inset is a plot of the series resistance, as a function of temperature, derived from this I-V data set.

Fig. 2: Cu-Kα X-ray diffraction data for B-C layers on SiC and Si substrates: (a) on SiC, (b) and (c) different depositions of B-C on Si. Major and minor B-C layer peaks that are common to growth on SiC and Si are marked by symbol ■ in (a).

Table Caption

Table I:  Bragg 2θ diffraction angles and interplanar spacings obtained from Cu-Kα x-ray diffraction by B-C layers on Si and SiC substrates. Also listed are the ICDD Powder Diffraction File (PDF) data for closely matching crystalline phases and associated crystal plane indices. Bold lettering indicates the closest agreement.

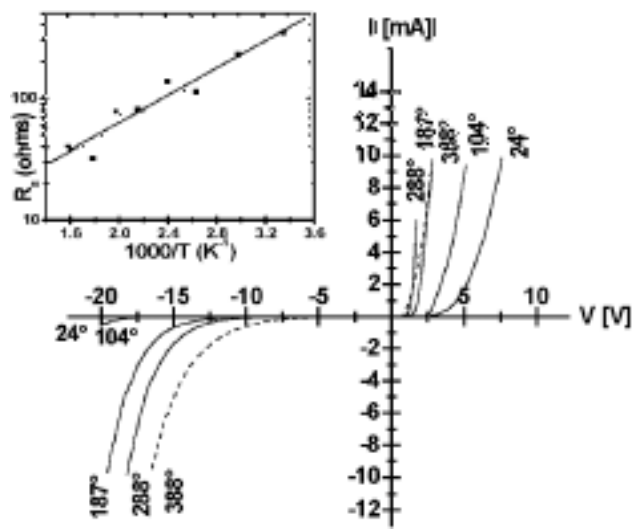

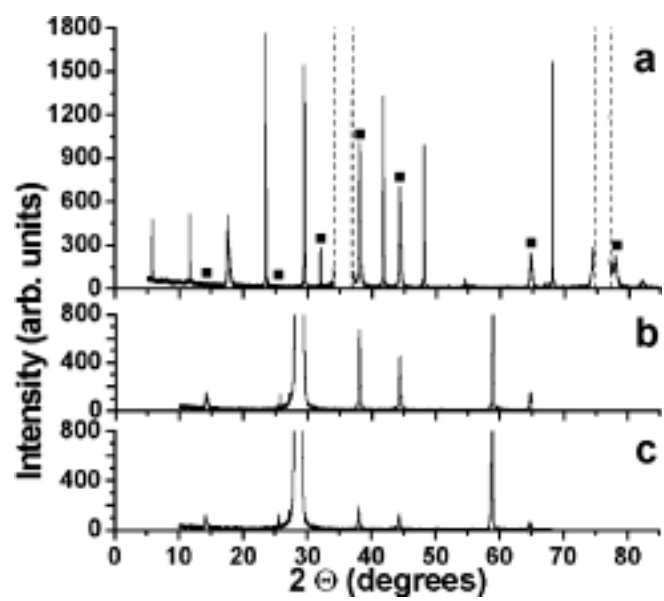

| 2θ (deg.) | Plane spacing (nm) | Phases (from ICDD PDF) | PDF spacing (nm) | PDF plane index |
|---|---|---|---|---|
| 14.24 | 0.614 | **Tetrag. B** or **B48B2C2** | 0.617 0.617 | (1 1 0) (1 1 0) |
| 25.31 | 0.352 | **Rhomb. B** | 0.352 | (1 1 0) |
| 28.70 | 0.311 | **B48B2C2** or Tetrag. B | 0.309 0.308 | (2 1 1) (2 2 0) |
| 29.60 | 0.302 | **Tetrag. B** | 0.308 | (2 2 0) |
| 32.21 | 0.278 | **Rhomb. B** or B48B2C2 or B12(BC2) | 0.277 0.276 0.281 | (3 3 1) (3 1 0) (1 1 0) |
| 34.21 | 0.262 | **Rhomb. B** or B48B2C2 | 0.261 0.264 | (-2 2 1) (2 2 1) |
| 38.28 | 0.235 | **Rhomb. B** or B48B2C2 | 0.236 0.238 | (-3 1 1) (1 1 2) |
| 44.49 | 0.204 | **Tetrag. B** | 0.206 | (3 3 0) |
| 64.60 | 0.144 | **B48B2C2** | 0.144 | (3 1 3) |
| 77.85 | 0.123 | **B48B2C2** or **Rhomb. B** or B41.1C4.45 | 0.123 0.123 0.122 | (5 5 0) (7 7 4) (4 0 1) |